# AIED's Unfinished Mission: Centering Agency and Motivation in the Age of Effortless Bypass

H. Chad Lane(✉)

University of Illinois, Urbana-Champaign, Champaign, IL 61820, USA
hclane@illinois.edu

**Abstract.** The widespread availability of general-purpose AI that can perform complex cognitive tasks threatens to undermine education at scale. This *effortless bypass dilemma* sharpens a challenge AIED has long engaged with but must now confront directly: ensuring learners choose effortful engagement when easier alternatives are available to complete learning tasks. In this paper, I argue that AIED's longstanding agenda of building more effective intelligent educational tools should continue, but with a renewed emphasis on the urgency of ensuring learners choose to engage authentically. Drawing on established motivational and learning theories, I outline five directions in which AIED can build on its existing strengths: supporting autonomy and agency, building learner resilience to metacognitive threats, designing for interest and relevance, amplifying process-based assessment, and empowering teachers. I then share four envisioned technologies that embody key features of this future and conclude by outlining how AIED must now evolve.

## 1 Introduction

### 1.1 The Changing Context of AIED Research

The threat of generative AI (GenAI) to undermine education has not gone unnoticed in the popular media. A review from the Brookings Institution concludes that the risks of GenAI to children's foundational learning, well-being, relationships, and privacy currently outweigh its educational benefits [9]. The authors call to "bend the arc of AI implementation toward educational experiences that help all children flourish academically, socially, and civically" (p. 10). Despite limited recognition of research demonstrating the positive effects of intelligent tutoring systems (pp. 45–46) for personalized learning, this extensive report largely ignores AIED and its over 35 years of research [14] and its significant convergence with what are precisely their stated goals.

The Brookings framing collapses a crucial distinction: general-purpose AI products adopted by students and schools without pedagogical design are not the same kind of educational technology AIED has spent decades building. When AI-based educational technologies are designed in ways that are consistent with learning sciences research and rigorously evaluated, as has been the norm in AIED, the evidence for effectiveness is not mixed at all. AIED has consistently produced positive results showing that AI-based tools approach learning gains comparable to human tutoring [37]. A sweeping "mixed benefits" verdict on "AI in education" risks burying that evidence under the much noisier record of unprincipled deployment in practice.

But, GenAI has undeniably changed the conditions under which AIED operates. Learners now have access to general-purpose AI tools that can perform academic tasks along the full cognitive pipeline, across domains, and to a high degree of quality. These tools can be used anonymously, relatively effortlessly, and outside any designed learning environment. Unchecked and unmonitored, it is now trivial to complete standard homework assignments (e.g., math and science problems, essays, programs) with simple prompts. I will refer to this challenge as the *effortless bypass dilemma*.

This shifting of effort to technology, *cognitive offloading*, is neither new nor unique to AI [25]. Calculators and the internet both offered easier paths for students who sought to outsource their homework. These tools could even be positioned to have value: for example, using a calculator to check one's work, or synthesizing several web pages to arrive at a well-reasoned conclusion. Similarly, while modern AI tools (e.g., ChatGPT) can be used to "bypass productive thinking," [18], I find it unsatisfying to simply limit their use since it harms only the learners who cannot access AI outside of school. More nuanced (and creative) approaches are needed that address the real underlying challenges. What can educators and researchers do to increase the chances that learners will choose to embrace the cognitive effort needed for effective learning? How can we establish the conditions that promote deep engagement and cultivate what Mark Lepper called "a desire to be taught" [20]? My central thesis is that (1) we should look to educational and motivational theory for guidance on how to reframe the future of AIED, and (2) that AIED has already been working on many of the relevant foundational issues for decades.

### 1.2 A New Status for Agency, Motivation, and Self-Regulation

The effortless bypass dilemma is, at its core, a motivation problem. We aspire for learners to understand the value of engaging cognitively, and to choose to do so for the right reasons. With this characterization, established motivational theories make precise predictions about this dilemma. Self-Determination Theory (SDT) [12] warns of an autonomy paradox: AI tools feel empowering but can shift the locus of causality from learner to tool, producing dependency rather than genuine agency. Achievement Goal Theory (AGT) [23] predicts who is most vulnerable: performance-oriented students have a rational incentive to bypass when the goal is limited to a product, while mastery-oriented students are more protected because the learning process itself is what they value. Self-Regulated Learning theory (SRL) [40] identifies the mechanism through which bypass causes the most damage: AI can short-circuit the forethought, monitoring, and self-reflection cycle through which learners develop independence. This has led to what researchers have called *metacognitive laziness* [15]. Finally, in the case of interest development [24], AI has the potential to interrupt the growth from early phases of interest to more advanced phases, when sustained self-generated motivation fuels continued re-engagement.

These frameworks converge on a common insight: the productive struggle that AI can eliminate is not an unfortunate byproduct of learning, it is the mechanism through which learning occurs. Emerging evidence confirms these predictions. A recent study found that unrestricted ChatGPT access showed substantially better practice performance but significantly worse exam scores [3]. Other recent work clarifies that providing learners with choices about AI use is not equivalent to supporting genuine agency [7], which requires the capacity to make and enact good learning decisions.

### 1.3 AIED Foundations

For the AIED community, factors underlying the effortless bypass dilemma are thankfully not novel challenges. The field boasts an extensive historical (and ongoing) focus on the foundations of motivation, agency, and self-regulation. The roots and values of AIED are no more clearly stated than John Self's pioneering vision of building "Systems that Care" which argued we should design tools to address learners' motivational, cognitive, and affective needs [32]. Du Boulay et al. enacted the vision with explicit integration of motivation, affect, and metacognition into an intelligent tutoring system [6]. For decades, Bull & Kay (and many others) have centered learner control and agency with their work on open learner models, which gave students genuine access and control over their own learning data while promoting self-regulatory skill development [8]. Aleven et al. modeled help-seeking as metacognitive behavior [1], essentially defining the problem space that GenAI now explodes (and gives learners immediate access to alternative tutors if they don't like using ours). Biswas et al. [4] demonstrated through Betty's Brain that positioning students as teachers keeps them cognitively active, a design that structurally resists bypass by reframing the learning task. As a last illustrative example, Baker et al. showed the risks of bypass and the draw learners feel when AI can help in seminal work on gaming the system (i.e., "help abuse") [2]. These contributions, among many others, provide the evidence that AIED has long recognized the richness of these challenges, but also offer a foundation on which the field can push ahead. What follows is a research agenda that draws on and extends this work and seeks to capture the guidance we can draw from foundational motivation and learning theories.

## 2 Revising the AIED Research Agenda

### 2.1 Supporting Autonomy and Agency

SDT identifies autonomy as the experience of ownership and control over one's actions. It is characterized as a basic psychological need that drives internal-ized motivation, along with SDT's other two pillars, competence and relatedness [12]. As noted above, modern AI creates a paradox here: tools that feel empowering can shift the locus of causality from learner to the tool, producing dependency rather than genuine self-direction. Brod [7] provides a useful framework for understanding what authentic agency actually requires in educational contexts: not merely the opportunity to make choices, but the capacity to make good decisions and enact them. This distinction matters for AIED and suggests the explicit consideration of what may happen outside of the scope of a system we build. It also has implications for AI literacy needs that have historically not been relevant to the popular "homework-helper" flavor of systems we have created.

AIED has made significant strides investigating agency and autonomy. Open learner models (uniquely from AIED) provide the ability for learners to inspect, challenge, and even negotiate the system's model of their knowledge [8]. This framing centers learner agency and extends naturally to the current challenges and common demands for interpretability and scrutability of AI systems. Emerging research questions might be: Can the learner have genuine input into what learning activities they undertake and why? Can they understand the rationale behind a given assignment? This would represent a shift from simple transparency of a learner model to co-construction of tasks and the design of a learning experience itself. Such levels of control have significant downstream motivational benefits [11]. How we model a learner's developing capacity for agency, how we scaffold it appropriately across ages and contexts, and how we distinguish genuine self-direction from choosing the path of least resistance are open research questions that the field is well positioned to pursue.

Critically, research certainly does not suggest that unbounded agency is ideal for learning. Taub et al. [36] found that students given a moderate degree of agency in a game-based learning environment achieved better learning outcomes than students with either no agency or high agency, while still maintaining interest and engagement. The implication for AIED is significant: calibrating the right amount of agency for a given learner, task, and moment is itself a design and modeling problem, and one that likely demands differentiation and high levels of adaptivity.

### 2.2 Building Learner Resilience to Metacognitive Threats

SRL theories tend to involve a cyclical process including forethought, performance monitoring, and self-reflection. As SRL skills mature, learners develop the capacity to manage their own learning [40]. Fan et al. [15] demonstrated that this cycle is vulnerable to GenAI: in their study, students using ChatGPT showed collapsed transitions between orientation, planning, monitoring, and evaluation, with metacognitive activity restructuring around the tool rather than around learning. Learners tended to offload not just cognitive work, but also SRL processes. Xu et al. [38] subsequently showed that embedding metacognitive prompts in AI interactions could restore these processes, thus offering a potential form of resilience to the longer term risks.

What happens when AI access is pulled? Liu et al. [21] enrich the picture with randomized experiments showing that brief AI-assisted practice not only reduces immediate engagement but carries over: participants who lost access to AI afterward solved fewer problems and gave up more often than participants who never had AI assistance. Critically, these persistence costs were concentrated among learners who used AI to obtain direct solutions; those who used AI for hints or clarifications showed no such impairments, a pattern that maps directly onto the productive-use versus bypass distinction and AIED's substantial history investigating help-seeking and scaffolding.

AIED has been building toward the challenge to promote and strengthen SRL skills for decades. Aleven et al.'s [1] work on help-seeking behavior in intelligent tutoring systems modeled how students make decisions about when to seek assistance and when to persist without. Such decisions map directly onto the bypass dilemma students now face with GenAI, but at a far greater scale and with reduced barriers. The field's extensive work on SRL scaffolding within tutoring systems provides a strong technical foundation [13,26]. A critical research frontier relates to the durability of SRL support: can AIED systems scaffold SRL skills that hold up when learners leave the scaffolded environment and encounter AI tools that offer no such support? Designing for resilience and to help learners internalize self-regulatory habits is the challenge that distinguishes this moment from previous work on metacognitive scaffolding.

### 2.3 Designing for Interest and Relevance

Renninger and Hidi's four-phase model of interest development describes a trajectory from triggered situational interest, through maintained situational interest, to emerging and eventually well-developed individual interest [24]. Each phase brings distinct characteristics and support needs, and the transition from situational to individual interest is where engagement becomes self-sustaining. Without the necessary interest triggers and early cognitive engagement that follows, bypass (via AI) threatens movement to more enduring forms of interest.

Interest (and its cousin, curiosity) are perhaps among the most undervalued constructs in AIED research. Despite their centrality to motivation and engagement, relatively little published work in the field has treated interest as a distinct design target. AIED researchers have only begun to address the question of how to deliberately detect, trigger, sustain, and develop learner interest [19,39]. This gap was understandable when engagement was required via compulsory work, but it now seems more important than ever as the AIED research community to cultivate and promote learner interest in the topics we hope they will learn.

GenAI creates a genuine opportunity here. AIED systems, more or less, have operated with relatively fixed content, generating problems from templates, and adapting difficulty or sequencing work. But, fundamentally reshaping learning tasks around individual learner interests has remained aspirational. GenAI's flexibility may bring such learning experiences into scope. For the first time, we may be able to dynamically adapt learning contexts to what a student cares about while preserving underlying learning objectives. Research is needed on how to model interest development computationally, how to design systems that trigger and sustain interest rather than merely measure engagement, and how interest-driven task design affects learners' willingness to persist through difficulty rather than bypass it.

### 2.4 Amplifying Process-Based Assessment

When assessment is limited to considering a learner's output, bypassing the work is incentivized. When assessment captures how students think and work, and that is used to formulate feedback, the focus shifts to process, which in AGT is consistent with mastery goals. This principle is well established in both motivation research and assessment research; it also takes on new urgency when AI can generate polished outputs indistinguishable from student work.

Unsurprisingly, AIED researchers have been building the infrastructure for process-based assessment for a long time. Nearly two decades ago, Shute et al. offered a "peek into the future" of AIED assessment that already envisioned continuous, unobstructed measurement woven into intelligent learning environments [34], a vision that has only become more urgent with GenAI. The corresponding notion of stealth assessment [35] involves the embedding of evaluations (invisibly) within learning activities, measuring competencies through behavior rather than distinct testing events. The entire field of educational data mining, which grew directly out of a series of AIED workshops, is built around making sense of process data, developing sophisticated methods for analyzing sequences of learner actions, help-seeking patterns, and more. [27].

The research challenge continues to be to make these tools accessible and practical for practitioners who urgently need alternatives to product-based evaluation. Can existing process analytics methods scale to the more open-ended interactions that GenAI creates? Can they reliably distinguish productive AI use from bypass? And can we enrich assessments of knowledge to also capture critical noncognitive skills such as agentic engagement, self-directed learning, and embrace of desirable difficulty? These markers of intrinsic motivation are ultimately the best defense against bypass and therefore have elevated importance in the future of AIED research.

### 2.5 Empowering Teachers with AI

The argument so far has centered the learner's personal decision to bypass, but that decision is never made in isolation. Teachers set the conditions under which those decisions happen: they calibrate difficulty, legitimize struggle, recognize productive versus unproductive disengagement, and build the relationships that sustain motivation over time. In other words, autonomy, mastery orientation, interest, and self-regulation are all heavily influenced by how teachers shape the learning environment. A research agenda focused on learner agency is therefore incomplete without a parallel commitment to teacher agency and empowerment. AI that deskills teachers, floods them with data, or displaces their pedagogical judgment undermines the very conditions on which learner motivation depends.

AIED is again on solid ground in thinking about a future of teaching with AI (or even, despite AI). On top of decades of research focusing on tutoring tactics, feedback, and scaffolding, AIED researchers have built a wide range of systems that support pedagogical decision-making in the classroom [16]. Recent work on AI-supported teacher tools and the vision of Human-AI complementarity in education builds on the idea that humans and AI systems bring unique strengths [17]. Such systems can help teachers monitor classroom activity and identify students who need attention in real time. The AIED research challenge is to extend this work so that teachers are empowered to orchestrate the kinds of learning environments that continue to support cognition, but also enable critical opportunities for agency, autonomy, and mastery. AI that amplifies teacher expertise and passion, and that helps implement teacher goals (rather than adding to their workload), is an obvious goal to set. How we design these partnerships, and ensure they remain grounded in what we know about effective teaching and learning, is an essential part of the future AIED research agenda.

## 3 What AIED Technologies Might Look Like

To illustrate this renewed vision for AIED, both in light of theoretical guidance from motivation and SRL science, as well as in the context of emerging AI capabilities, I share several feasible use cases. These are intended to be feasible to build (and may very well already exist!), but also informed by the substantial relevant work done in AIED over four decades.

### 3.1 Negotiated Learning Tasks

What if, instead of receiving a pre-defined assignment, a learner engaged in a conversation with AI about what work to do? In this envisioned system, the AI knows the teacher's learning objectives and the student's current state (knowledge, affect, dispositions), and draws out the student's interests, background, and ideas. Together they negotiate a task that meets curricular requirements while being personally meaningful. Following the principles of interest development theory, the system could supplement the necessary support structure for reaching advanced states of interest. Tasks could build on prior work, even over months and years. A teacher might specify that students need to demonstrate particular knowledge components. If the focus is on coding, one student might design a puzzle game, another a sports simulation, with both meeting the same objectives through work they helped shape. The AI system ensures that the work will tap into curricular goals and tap the targeted knowledge.

This approach directly enacts several theoretical principles: it supports genuine autonomy by giving learners meaningful input into their own learning (SDT), it shifts goal structures away from "complete the assignment" toward personally relevant mastery (AGT), and it operationalizes interest development by making the elicitation of learner interests the starting point of task design rather than an afterthought. The teacher's role shifts from creating individual problems to defining the space within which learning occurs and articulating learning objectives. This becomes an AI-teacher partnership in which the AI ensures curricular alignment while the student drives content and context. The negotiation itself is a metacognitive exercise: the student must articulate what they know, what they want to learn, and why it matters. Key questions for AIED research include how to balance learner agency with alignment to consistent learning goals, what distinguishes productive negotiation from task avoidance, and whether task ownership through negotiation increases willingness to persist through difficulty rather than bypass it.

### 3.2 Bypass-Aware Tutors

Baker's seminal work on gaming the system [2] demonstrated that students' unproductive behaviors within tutoring systems could be reliably detected and that specific design features made gaming more or less likely. A bypass-aware tutor extends this paradigm to the GenAI era. Rather than attempting to prevent bypass through restriction, which SDT predicts will produce backlash, this envisioned system detects engagement patterns and responds with adaptive metacognitive scaffolding. The system would learn to identify patterns associated with outside (efficient and effortless) "help". When bypass patterns emerge (simple ones being sudden jumps in solution quality, large text pastes, surface-level interaction), the system might gently prompt the learner to walk through the reasoning and explain it. It could also offer a similar problem, then invite the learner to try and solve it together (inside the tutor rather than with an external tool).

No detection scheme will ever be without vulnerabilities, of course. A determined student can simply inspect a large language model's (LLM) "thinking" and then retype the steps with plausible pacing. Distinguishing sophisticated bypass like this from genuine competence is perhaps a bridge too far without more invasive monitoring. But this is precisely the point: purely technical bypass detection is an arms race AIED cannot win on its own, which is why the much harder challenge of cultivating learners who do not want to bypass in the first place is now urgent.

Supporting learners who may be attempting to bypass directly addresses the metacognitive resilience challenge: it targets the collapsed self-regulation that Fan et al. documented [15] by re-engaging the forethought and self-reflection phases of Zimmerman's SRL cycle at precisely the moments when they are most likely to be bypassed. Over time, the system builds a longitudinal profile that could even be made visible to the learner (in the open learner model tradition). Visually displaying learner behaviors, such as when they engage deeply in cognitive effort versus when they offload, and how their unassisted performance compares to their assisted work, could have significant impact on some learners (perhaps even as a section in a student's report card). Making this gap visible supports accurate self-assessment calibration and addresses the learning vs. performance distinction (the idea that learners often confuse good performance with effective learning). The tone would need to be diagnostic, not accusatory:

a cognitive mirror that builds self-awareness rather than enforcing compliance. Research is needed on what interaction patterns might reliably distinguish productive AI use from bypass, and whether making engagement patterns visible to learners improves their self-regulation over time.

### 3.3 AI-Generated Game-Based Learning Environments

For this (imagined) use case, we start with an emergent capability that introduces new possibilities for education (much like mobile devices enabled advances in outdoor education). Specifically, we consider Google DeepMind's Genie 3 which can generate interactive, explorable 3D environments from a text prompt in real time.[1] These are navigable environments with consistent physics and the ability to modify the world dynamically through natural language. The educational implications, which Google has not been shy about pointing out, are significant given a vision driven by autonomy, agency, and interest laid out here. With such a tool, a student could simply describe a setting and step into it moments later. Advances like this are impossible to ignore when envisioning how AIED moves ahead. Pre-authored educational games have always been constrained by the cost and rigidity of fixed content, which is often a strength (especially from storytelling and design-based frames), but AI-generated worlds bring something entirely different that captures new possibilities.

Games are among the few contexts where people voluntarily seek out difficulty and persist through failure, and this is not coincidental. Games naturally incorporate several motivational principles that the bypass dilemma demands: challenge calibration keeps players in a productive zone that mirrors the zone of proximal development, mastery goal structures arise organically when the experience itself is the reward (AGT), and intrinsic interest is sustained through narrative, agency, and meaningful choice (SDT, interest). In a game, using cheat codes can undermine the experience and sense of satisfaction, suggesting bypass might be positioned as motivationally self-defeating rather than merely educationally harmful.

AI-driven non-player characters could serve pedagogical roles, just as they have in Crystal Island for nearly two decades [29], and engage in far more impactful conversations than historically possible, as collaborative partners, characters who need the player's help to understand something (embedding the teachable agent paradigm), or Socratic questioners who push the player to articulate their reasoning. Recent work from the NSF EngageAI Institute points to what is now within reach. Their *SceneCraft* tool uses LLMs to generate interactive narrative scenes (dialogue, characters, branching paths) from natural language prompts, with teachers in the authoring loop [28]. This lowers a long-standing barrier in narrative-centered learning, namely the cost and rigidity of hand-authored content, and makes it feasible to tune stories around individual learner interests and curricular goals, exactly the conditions interest development theory predicts will sustain engagement.

---
[1] https://deepmind.google/models/genie/.

Gameplay also generates rich process data for assessment: problem-solving strategies, persistence patterns, exploration behavior, and help-seeking. Research questions for AIED include how to ensure AI-generated challenges maintain educational rigor, whether the motivational structures of gameplay transfer to non-game contexts, and what design features most effectively sustain interest across developmental phases.

### 3.4 Giving Teachers "Superpowers"

This final use case explores teacher-AI partnerships in the classroom, and specifically how AI can be used to amplify teacher agency and control. AI-driven classroom dashboards have been a mainstay of AIED and learning analytics for decades, surfacing indicators of student progress, struggle, and engagement to teachers [10]. Positioning AI in more of a direct partnership role, Holstein et al. pushed this notion further by pursuing a mixed-reality tool for teacher augmentation called *Lumilo*. By offering data-driven, context-relevant insights to teachers, this work demonstrated how just-in-time teacher support can produce measurable learning gains in the classroom [17]. Seminal work like this revealed that fine-grained interaction data collected by AIED systems (such as intelligent tutors) could be leveraged to offer actionable insights for teachers.

But what if teacher-facing AI tools acted more like a conversational partner? With access to every student's interaction data and the ability to assess that data in terms of knowledge components, engagement levels, and affect, a teacher would be able to interrogate classroom data efficiently and in targeted ways. Imagine a teacher asking in natural language: "How did the class do on assignment 3 yesterday?", "Did my second period persist through those debugging challenges on Friday?", or "What kind of help did Devon get from the agent on his homework?" The AI answers conversationally, grounds its claims in the actual logs, and flags its own uncertainty (exhibiting "epistemic humility"). The system might also ask the teacher for help in refining its own learner model estimate for a specific student (e.g., "Was Frank distracted in class? I see some unexpected variance in his engagement levels this week.").

To fully empower a teacher, the envisioned system would need to extend to controlling how AI behaves in the classroom, such as with helping teachers implement adjustments to instruction such as by controlling ITS policies (e.g., "Prompt Maya to self-explain during her homework exercises") or adapt assignments (e.g., "Add some easy warm-up assignments on nested loops today"). Much in the spirit of Lumilo, this envisioned system recognizes the complementarity of AI's access to raw data with that of the teacher's instincts, honed by years of practice and classroom experiences. Emerging work on conversational interfaces to data and LLM-supported data storytelling [31] suggests that this vision is now within reach.[2]

[2] In the NSF INVITE AI Institute, we are developing a prototype of this envisioned system called *Project Keating*.

# 4 AIED 2.0

## 4.1 What Changes in Practice

The research agenda outlined in this paper has methodological implications. Studies that eliminate the possibility of bypass may still have value, but will carry a higher cost in ecological validity. If we want to understand how learners behave when bypass is an option, we must study environments where it is present. This calls for more longitudinal work tracking the effects of sustained AI use on motivation and self-regulation over time, mixed-methods approaches that capture not just whether students learn but why they engage or disengage, and new evaluation metrics.

We should consistently track and value (as researchers) voluntary engagement measures alongside learning gains. We see again that this is not a new question for AIED: Shute [33] showed that learners' hint-asking patterns carried diagnostic information about learning, with the meaning of those patterns depending on error rates and timing rather than on help-seeking alone. If a system produces learning gains only under compulsory use, it is worth asking whether that constitutes success in a world where compulsory use with deep engagement is increasingly unenforceable. The field also needs deeper interdisciplinary partnerships, not just citing motivation theories but co-designing research with motivation scientists, social scientists, and emotion psychologists who bring methodological traditions AIED has not always drawn on.

AI literacy itself has seen rapid growth in attention in AIED recently, and deserves such a rise in our attention, both for measurement and as a targeted domain for learning. Without a read on what learners actually know about AI's capabilities and limits, behavioral data in a post-GPT world is hard to interpret: a student making heavy use of ChatGPT on a writing task may be exhibiting productive, metacognitively aware tool use or an abdication of thinking, and we miss this from logs alone. AIED therefore has a double stake in AI literacy, as a construct we need to measure to make sense of engagement data and as a capability our systems should help learners develop.

## 4.2 What Changes in Vision

AIED has long defined its success in terms of building more effective educational AI tools for teaching, learning, and research. Of course, we must continue to embrace these goals and recognize that knowledge acquisition is and will always be central to education. But we must also continue to ask if effectiveness measured solely by learning outcomes remains the one that "really counts". This challenge is obviously inherited from AIED's context (e.g., standardized testing), but AIED researchers have also pushed back on this unwanted convention, with significant work on social and emotional learning [5] and embodiment [30], to name a few examples.

Here, I have argued that the extent to which learners choose to engage authentically matters now more than ever. Whether learners develop the self-regulatory capacity to manage their own (independent) AI use, and whether the systems we build cultivate motivation rather than merely delivering content, represent some fundamentally different outcomes AIED historically pursues. The envisioned technologies explored in Sect. 3 are certainly not intended to simply replace effective tutoring systems, but they do represent attempts to take AIED strengths, combine them with emerging capabilities, and meet a new set of goals focused on motivational and self-regulatory outcomes. They are examples of what becomes possible when previously "nice-to-haves" are elevated to design imperatives alongside learning outcomes. AIED's existing traditions in learner modeling, metacognitive scaffolding, pedagogical agents, and process-based assessment provide exactly the foundation needed. The pieces are largely in place; the challenge is to redirect them.

We must also recognize that AIED inherits the institutional contexts it works within. No matter how sophisticated a process-based assessment may be, for example, its impact is blunted when institutions continue to credential only outputs. Teachers adopting bypass-aware, agency-centered designs end up working against the grain of transcripts and grading structures that reward polished final artifacts over the learning that produced them. AIED 2.0 must therefore acknowledge institutional adaptation as part of its landscape: diplomas and transcripts will need to reflect not only what learners know but how they came to know it, including the effort and growth in learning capability that process data can increasingly document.

### 4.3 What Changes in How We Engage

The proposals in this vision are also an undeniable example of AIED's new reality. Unprecedented investments in (industry) AI have produced an extraordinary pace of change. The downstream effects on schooling, teaching, and the relationships learners now have with technology are critical inputs into how we plan as a research community. We must understand how learners actually interact with rapidly evolving tools. AIED cannot afford to assume stability in the learning environments in which learners (or we) work. If the field conducts research based on assumptions about what AI can and cannot do, those assumptions may be obsolete before a paper is published ("What if that tutoring system was built on ChatGPT version N+1?"). Staying relevant requires active engagement beyond our own community, and being a part of conversations focused on AI literacy, school policies towards AI, and tracking how teachers are integrating off-the-shelf tools they find or even make.

We should be attending industry AI conferences to understand where the technology is heading. We should be asking students directly how they use AI, what they do with it, and what it means to them. To me, this means more than reading the latest report or survey; when we are in schools, around teachers and students, we should actively talk about it and learn from these audiences. We should be in close dialogue with teachers navigating these changes in real time, and with policymakers making consequential decisions about AI in schools.

We could all draw inspiration from Rich Mayer, a highly-cited cognitive psychologist whose body of work is foundational to many AIED systems and who has served over four decades on his local school board. He documents his many lessons learned in his 2011 book *How Not to Be a Terrible School Board Member* [22]. This kind of direct, sustained engagement with the administrative machinery of schooling has historically been the exception in our community. It may need to become closer to the norm: the credentialing structures, district policies, and procurement decisions that determine whether our research actually lands on learners are made in rooms AIED researchers rarely sit in.

Adapting is not optional if we want AIED research to matter in practice and on the front lines of education. The field has the expertise, the methods, and the track record to lead on the most consequential educational technology challenge of a generation. But achieving such a status and recognition worldwide will require us to revisit our own history and goals, perhaps around agency and motivation, or maybe in some way we haven't quite figured out yet.

**Acknowledgments.** I wish to thank Suma Bhat, Val Shute, Kurt VanLehn, Diego Zapata-Rivera, and Shan Zhang, who provided comments on an earlier draft of this paper. This material is partially based upon work supported by the U.S. National Science Foundation and Institute of Education Sciences (INVITE AI Institute) under Award No. 2229612.

**Disclosure of Interests.** The author has no competing interests to declare that are relevant to the content of this article.